\documentclass{aa}
\usepackage{graphics}
\begin{document}
 
\title{A new population of soft X-ray weak quasars}
 
\author{G. Risaliti\inst{1}, A. Marconi\inst{2}, R. Maiolino\inst{2},
M. Salvati\inst{2}, P. Severgnini\inst{1}
}
 
\institute{
Dipartimento di Astronomia e Scienza dello Spazio,
Universit\`a di Firenze, Largo E. Fermi 5, I--50125 Firenze, Italy
\and
Osservatorio Astrofisico di Arcetri, Largo E. Fermi 5,
I--50125 Firenze, Italy
}
\authorrunning{Risaliti et al.}
\titlerunning{X-ray obscured quasars}
\offprints{G. Risaliti}
 
\date{Received / Accepted}

\abstract{
Quasars selected in optical surveys by means of their
blue or UV excess are known to be strong emitters in
the X-rays, except for Broad Absorption Line
(BAL) objects. In this paper we study the X-ray emission of quasars selected 
through their emission line spectrum rather than their blue excess. X-ray
data are obtained by cross-correlating two optical samples (the Hamburg
Survey Catalogue and the Palomar Transit Catalogue) with the WGA catalogue
of ROSAT observations. We find that a significant fraction of objects are
strongly underluminous in the X-rays. We discuss the physical nature of
these sources and we propose an interpretation for their optical and X-ray
properties based on a dust-poor and gas-rich absorber. Our results suggest
the existence of a population of AGNs having a type 1 optical spectrum
and a type 2 X-ray emission. 
}
	
\maketitle
	
\section{Introduction}
The broad band properties of quasars have been investigated in 
recent years through multiwavelength observations of several quasar
samples: the mean Spectral Energy Distribution (SED) of quasars was
estimated for a sample of optically selected, 
X--ray bright quasars by Elvis et al. (1994);
the X-ray properties of optically selected quasars were determined
for several large samples by using the Einstein all--sky survey (Avni \&
Tananbaum 1986, Wilkes et al. 1994) and the ROSAT all--sky survey (Green
et al. 1995, Brinkmann et al. 1998, Yuan et al. 1999). The results of
all these works were
essentially that quasars emit a significant fraction of their bolometric
luminosity (several \% or more) in the X--ray band. An indicator widely used
in the literature, in order to measure the X--ray emission
relative to the optical, is the ``Optical to X index'' $\alpha_{OX}$
defined as:

\begin{equation}
\alpha_{OX}=-\frac{\rm{log}(L(\rm{2500 \AA})/ L(\rm{2 keV}))}{\rm{log}(\nu({\rm 2500
\AA})/ \nu(\rm{2 keV}))}
\end{equation}

\noindent where L and $\nu$ are the monochromatic luminosity and the
rest--frame frequency.
The above authors find on average $\alpha_{OX} \sim $ 1.5, with a
dispersion $\sigma(\alpha_{OX}) \sim 0.2$. Moreover, this index is found
to increase slightly with redshift and luminosity. Together with these
``normal'' quasars, a minority of objects is found to be very weak in
the X rays ($\alpha_{OX} \geq 1.8$):
 many of these sources are Broad Absorption Line quasars
(BAL), while others are optically normal objects\footnote{In this paper ``normal
quasars'' refers to objects selected with standard color techniques. Their mean SED 
is derived from Elvis et al. (1994), normalizing the X-ray emission
in order to have $\alpha_{OX} = $ 1.5. The optical line spectrum is derived from
Francis et al. (1991).}. 

As one can easily estimate from Fig. 1, X-ray weak sources are a few
per cent of the entire population of quasars.

Recent works suggest that the origin of the
X--ray weakness of BALs is absorption rather than an intrinsically weak
emission (Brandt et al. 2000). At the same time, BALs have no optical
reddening, hence their optical colours are those typical of normal
quasars: for this reason they have been detected by ``standard'' surveys
based on U and B colour excesses.                                    

Our search for a new population of soft-X-ray weak 
quasars is motivated by several
results that suggest the existence of objects other than BALs with a mismatch
between optical and X-ray classificationi, and/or optical colours redder than
normal quasars:

\begin{itemize}
\item The optical identifications of the HELLAS sources (hard X--ray
selected) indicate that some objects with very hard X--ray spectra are
type 1 AGNs in the optical (Fiore et al. 1999).
\item The optical and infrared photometry of a sample of flat spectrum
radio quasars shows that a significant fraction of radio--selected
objects have redder colours than ``normal'' quasars (Webster et al.
1995). A similar conclusion, even if based on a much smaller sample, has
been reached by Kim \& Elvis (1998) who searched for red quasars among
soft X--ray selected sources.
\item An X--ray study of a sample of very luminous, high redshift
quasars performed by the ASCA satellite revealed the existence of
several objects with heavy obscuration in the 2-10 keV band (Reeves et
al. 1997).
\item A comparison between the absorbing column density N$_H$, measured
through X--ray analysis, and the optical reddening E$_{B-V}$, measured
through the Balmer decrement, performed on a sample of local Seyfert
galaxies (Maiolino et al. 2001a), shows that for a large fraction of AGNs
the E$_{B-V}$/N$_H$ ratio is much lower than Galactic.
The same conclusion is also suggested by the 
comparison between the X--ray and mid-IR emission of AGN--dominated
Luminous Infrared Galaxies (Risaliti et al. 2000).
\end{itemize}

The basic idea of our work is that BAL quasars could be the tail of a
population 
of X-ray weak quasars. In this view BALs would be extreme objects because,
 despite of  their strong absorption in the X rays, 
their optical colours are typical of normal (i.e. blue) quasars.
Between the two
extremes of ``normal'' and BAL quasars, a population of objects
could exist, with a strong X--ray absorption but a low optical
reddening, enough to change slightly the optical colours (and, therefore,
to exclude the sources from colour--based surveys) but the associated extinction
would not be enough to obscure completely
the optical broad lines. According to this scheme, this new class of
objects should be classified type 1 AGNs if observed in the optical, but absorbed
(type 2--like) if observed in the X--rays. 
In the following we will refer with ``blue quasars'' to colour-selected quasars,
and with ``red quasars'' to the new population described above.          

The best way to search for red quasars in the optical is by means of spectroscopic 
selection. The basis of the work described here is a cross-correlation of 
grism-selected samples of quasars with optical and X-ray surveys.
We studied the optical colours and the X-ray to optical flux ratio of these objects
and compared them with a sample
of blue quasars. 

\begin{figure}
\centerline{\resizebox{\hsize}{!}
{\includegraphics{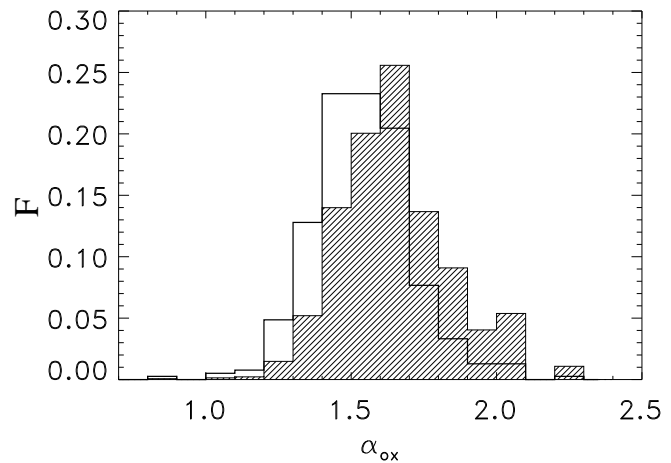}}}
\caption{\label{fig:alphaox}
\footnotesize{Open histogram: $\alpha_{OX}$ distribution
for the sample of optically selected quasars detected by ROSAT. Shaded histogram:
best estimate distribution taking into account also non--detections
(from Yuan et al. 1999).}}
\end{figure}

The paper is organized as follows: in Sect. 2 we analyze the selection
criteria adopted for the composition of the most studied quasar samples
and we define the characteristics of the quasar population that could
have been missed, then  we describe our
search for X--ray weak quasars, performed by cross--correlating spectroscopically
selected quasar samples with the WGACAT catalogue of ROSAT observations
(White et al. 1995). In Sect. 3 we present our results and in Sect. 4 we
discuss their
physical interpretation. In Sect. 5 we analyze the consequences of our
study for several relevant arguments concerning AGNs and the X--ray
background, and the future work in order to
define the properties of this new population of objects.

Throughout this paper we adopt the cosmological parameters H$_0 = 50$ km
s$^{-1}$ Mpc$^{-1}$ and q$_0=0.5$.

\section{Sample selection and X-ray data}

\subsection{Colour-based selection}
 
The optical emission of ``classical'' quasars is characterized
by two major features: an
optical--UV excess in the continuum and several strong, broad 
emission lines. The technique used in most quasar surveys
consisted in an automatic selection
of a reasonably large sample of quasar candidates, followed by spectroscopic
observations. Since quasars are rare objects and the spectroscopy is very 
time consuming, an efficient primary selection is needed. For this  
reason the color selection used in most surveys is very stringent: for
example, in the PG survey a color U-B $< -0.44$ is required and, although
this is a very blue color, the efficiency of the selection is less than 10\%,
the vast majority of candidates being blue stars or white dwarfs.

This kind of selection automatically excludes objects with intrinsically redder
spectra or with some absorption along the line of sight: for example,
assuming a galactic dust--to--gas ratio, a typical quasar at z=1 with a very blue
continuum would be excluded by the PG selection criterion if it is obscured
by a rest frame column density higher than $\sim 2 \times 10^{21}$ cm$^{-2}$.

\subsection{Broad line-based selection}

The considerations made above suggest that
 the best way to search in the optical for red or moderately
absorbed quasars is by means of a spectroscopic selection. The technique used
in spectroscopic surveys is usually a two-step selection, with a first step
at low resolution.

\begin{figure}
\centerline{\resizebox{\hsize}{!}
{\includegraphics{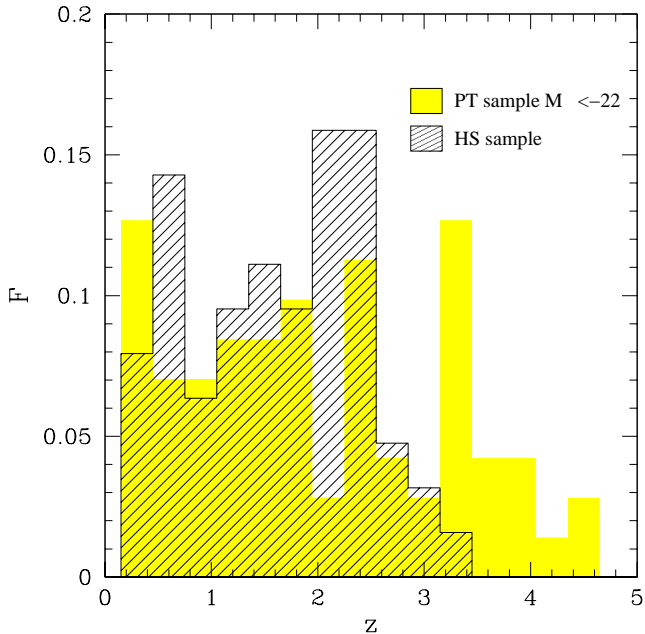}}}
\caption{\label{fig:zdist}
\footnotesize{Redshift distribution for the PT and HS samples.}}
\end{figure}

We have studied two spectroscopic samples of quasars.
The first one is the Palomar Transit Grism
sample (PT, Schneider et al. 1994), composed by $\sim$ 1000 objects selected
only by means of emission line criteria, without any requirement about
the continuum colours. Data available for these sources are the
redshift, an r4 magnitude and the flux, equivalent width and physical
width of the line used for the selection. The search technique was
optimized in order to discover many high--redshift quasars, nevertheless
most of the sources are at low or intermediate redshift. The redshift
distribution of the sample is plotted in Fig. \ref{fig:zdist}.

The second sample we used in our work is the Hamburg Quasar Sample (HS,
Hagen et al. 1995, Engels et al. 1998, Hagen et al. 1999).
In this case a first selection of the objects was performed by
using a color criterion (see Hagen et al. 1999 for details),
while a second filter consisted in the
requirement of broad emission lines in the observed spectrum. Even if a
colour selection is still present in this sample, the requirements on
the colours are less stringent than in ``classical'' quasar samples,
like the PG, and therefore many quasars with slightly redder
colours can be included. 

\subsection{Cross-correlation with the X-ray catalogue.}

The X-ray catalogue that we used for the cross--correlation with the
optical samples is the WGACAT, obtained by analyzing all the fields
observed in pointing mode by the ROSAT PSPC instrument.

We found that 180 sources of the PT and 85 of the HS are in the WGACAT
fields. 
Among the 180 PT sources, we made a further selection in order
to exclude objects with apparent magnitude E$>$18 and absolute
O magnitude M$_O >$ -22\footnote{E and O magnitudes are obtained from the
APM digital catalogue.}. A luminosity cut is required to avoid
contamination from the host galaxy. The optical brightness cut is due to our 
interest in X-ray weak objects: we therefore exclude the sources
that would be undetected in the WGACAT even if their optical-to-X ratio
were normal: indeed, in these cases we would only obtain X-ray upper limits
that are useless to constrain the X-ray properties of these sources.
The brightness cut could bias the sample in favor of objects with large
$\alpha_{OX}$. However, our control sample is the PG, with an even stronger brightness 
cut. So the ``differential'' effect cannot be spurious.

The final sample contains 30 PT quasars.

In the following we compare the X--ray properties of these
sources, relative to their optical emission, with the analogous
properties of the PG quasar sample. The PG sample is used as reference
both because it is the best studied quasar sample and because it is
selected according to the ``historical'' definition of quasars (i.e.
compact objects with high luminosity and a strong UV excess).

Among the 30 PT sources, only 9 are detected in the WGACAT, while for
the remaining 21 only an upper limit is available. Among the HS sources,
that are on average brighter than the PT sources, about half are detected. 
The upper limits have
been estimated at the 90\% confidence level, taking into accout the
vignetting and the PSF of the ROSAT PSPC instrument. Further details
are gven in the Appendix.

\subsection{Definition of an optical--to--X index}

The simplest way to compare the samples would be that of
calculating the $\alpha_{OX}$ index for line-selected sources, as defined in Eq. 1,
and following the prescriptions given in previous works (Green et
al. 1995, Wilkes et al. 1994). 
On the other hand, the calculation of $\alpha_{OX}$
poses several problems in the extrapolations needed: in particular, the
monochromatic flux at E=2 keV is strongly dependent not only on the
WGACAT count rate (that is the only information directly available), but
also on the spectral properties of the source. The standard method
consists in assuming a powerlaw spectrum with spectral index $\sim$ 1-1.5,
that is typical for quasars in the ROSAT energy band. However, we
suspect that our sources are much more absorbed in the X rays than
normal quasars: this would imply a possible photoelectric cutoff and/or
a much flatter spectral index. 

An analogous problem arises in the optical, even if in this case 
the errors should be lower.

For these reasons we define a new optical--to--X index, by using directly 
the available data: for the HS sample we use the B magnitude for the
optical and the WGACAT count rates for the X rays, while for the PT
sample we use the O and E magnitudes obtained from the APM sky
catalogue 
and the WGACAT count rates.

In order to perform a homogeneous comparison, we took the PG
quasars in the WGACAT ($\sim 70$) and we defined the count rates and the upper limits 
in the same way
as for the HS and PT samples.
The B magnitudes are directly available
in the literature (Schmidt \& Green 1983),
while the O and E magnitudes
were calculated from the B and V magnitudes 
assuming the quasar template of Francis et al. (1991). We note that
this extrapolation is much safer than those discussed above, because these
bands are much closer to B and V than 2500 \AA.

The optical--X index used in this work is defined as follows:
\begin{equation}
I_{OX}={\rm log} \frac{10^{(20-m)/2.5}}{{\phi(\rm cts~s}^{-1})}
\end{equation}
where m is the optical magnitude and $\phi$ the ROSAT count rate.
Since the optical magnitude does not always refer to the same band, we define a different
I$_{OX}$ for each band and compare only same-band indices. 
Finally, we note that in our
definition we do not take into account the K correction, even if
many of our sources are at intermediate or high redshift. We discuss
this point in detail in the next Section.

\section{X-ray properties of emission line selected AGNs}

In order to understand the physical properties of the selected objects,
we divide our sources in two
groups, to be studied separately: the HS sample and the PT sample.
This division is necessaray since the two parent samples have different
redshift distributions (Fig. 2) and selection criteria. 

\subsection{Analysis of the HS sample}

\begin{figure*}[!]
\centerline{\resizebox{\hsize}{!}
{\includegraphics{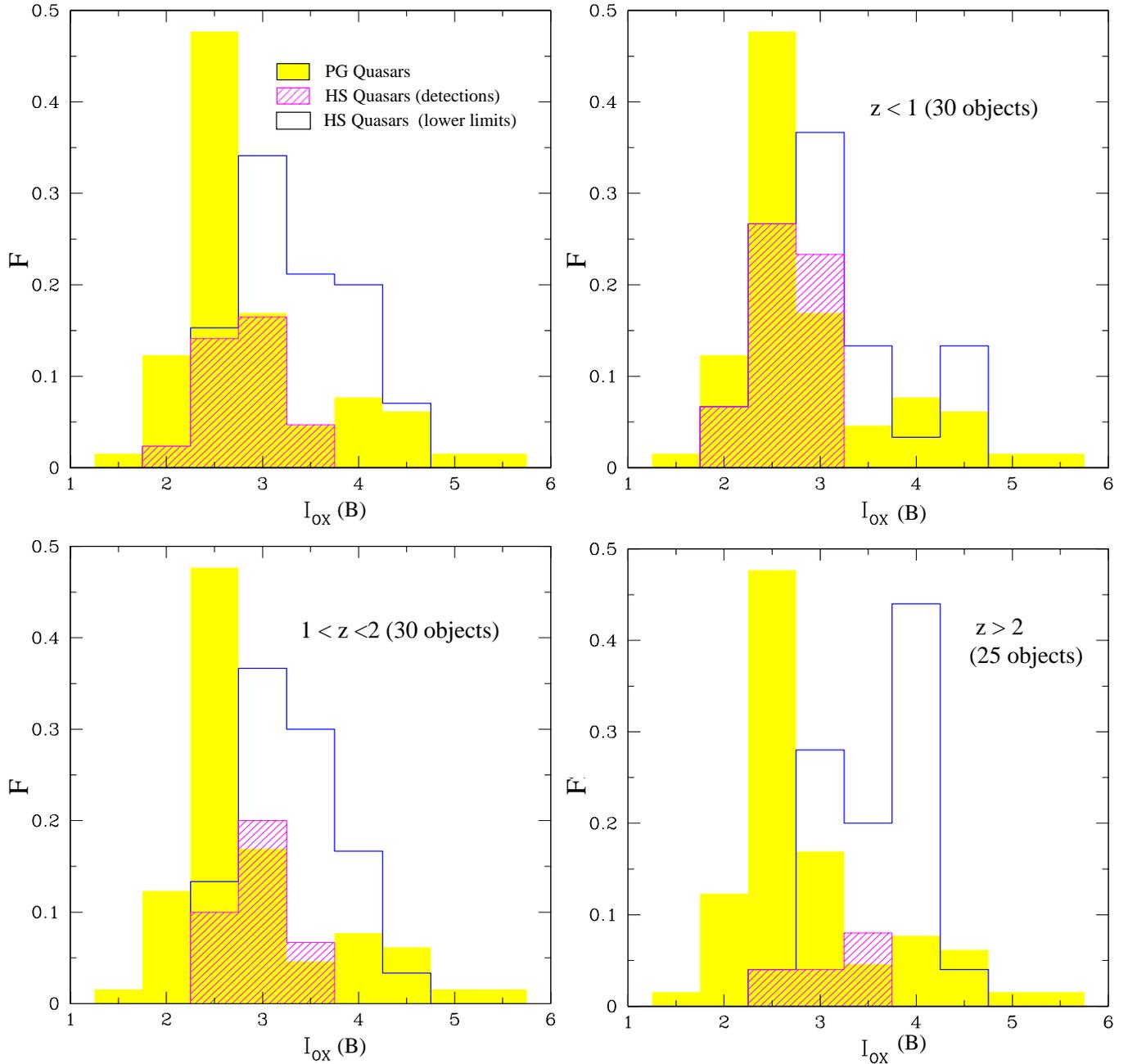}}}
\caption{\label{fig:hsiox}
\footnotesize{I$_{OX}$ distribution of the HS sample for
the whole sample (panel a) and for three redshift intervals. Lower limits
on I$_{OX}$ are derived from the upper limits on X-ray count rates.}}
\end{figure*} 

In Fig. \ref{fig:hsiox}
 we plot the I$_{OX}$ distribution for the entire HS sample
(Fig. \ref{fig:hsiox}a) and for three redshift intervals. We note
that no object in the HS sample is at
redshift z$<0.1$ and that high redshift sources are on average weaker in
the X-rays than intermediate redshift ones. A dependence of the
$\alpha_{OX}$ index (as defined in Eq. 1) with the redshift is suggested by
several statistical studies of large quasar samples, as reminded in the
Introduction. For this reason the comparison of the high--redshift
HS objects with the PG comparison sample could be not strictly correct, because 
the latter is composed mostly of low--redshift
sources. However, the redshift dependence that has been proposed in the
literature (a factor $\sim 3$ between z=0 and z $>$1, Yuan et al. 1999)
is much weaker than the effect found in the HS sample,
where the X--ray emission of a significant fraction of objects
is more than one order of magnitude lower,
with respect to the optical, than in normal quasars. 

Since the HS is a flux--limited sample, the redshift
dependence of the I$_{OX}$ distribution could be due either to a real
dependence on evolution or to a luminosity effect. This issue will be
discussed further in Sect. 4.

\subsection{Considerations on the K correction}
For logarithmic indicators, as
our I$_{OX}$ index, the K correction is $\sim (\gamma_1-\gamma_2){\rm
log(1+z)}$, where $\gamma_1$ and $\gamma_2$ are the spectral index
of the X--ray and optical spectrum respectively (assuming a powerlaw
spectrum, F$_\nu \propto \nu^{-\gamma}$).
Assuming a standard quasar spectrum, with $\gamma_1=2$ and
$\gamma_2=1.5$, the K correction is 0.15 at z=1 and 0.24 at z=2, much lower than
the observed effect.

It is well known, for example from
the analysis of the X--ray emission of BAL quasars (e.g. Gallagher et al. 1999),
that X--ray weak
quasars have on average flatter X--ray spectra than normal quasars. This
is also found in low-luminosity AGNs: if the complex 2-10 keV spectra of
heavily absorbed Seyfert 2s is fitted with a simple powerlaw, the photon
index varies between $\sim 0.5$ and $\sim 1.5$ (see for example Turner
et al. 1997 and Maiolino et al. 1998).
Therefore, the X--ray spectral index of our objects could be
significantly lower than the canonical value of 2, 
and this effect goes in the direction of lowering and perhaps inverting
the k--correction for I$_{OX}$. 

We conclude that a physical effect different from the K correction
is responsible for the
unusually high values of I$_{OX}$.

\subsection{Analysis of the PT sample}

\begin{figure}[!]
\centerline{\resizebox{\hsize}{!}
{\includegraphics{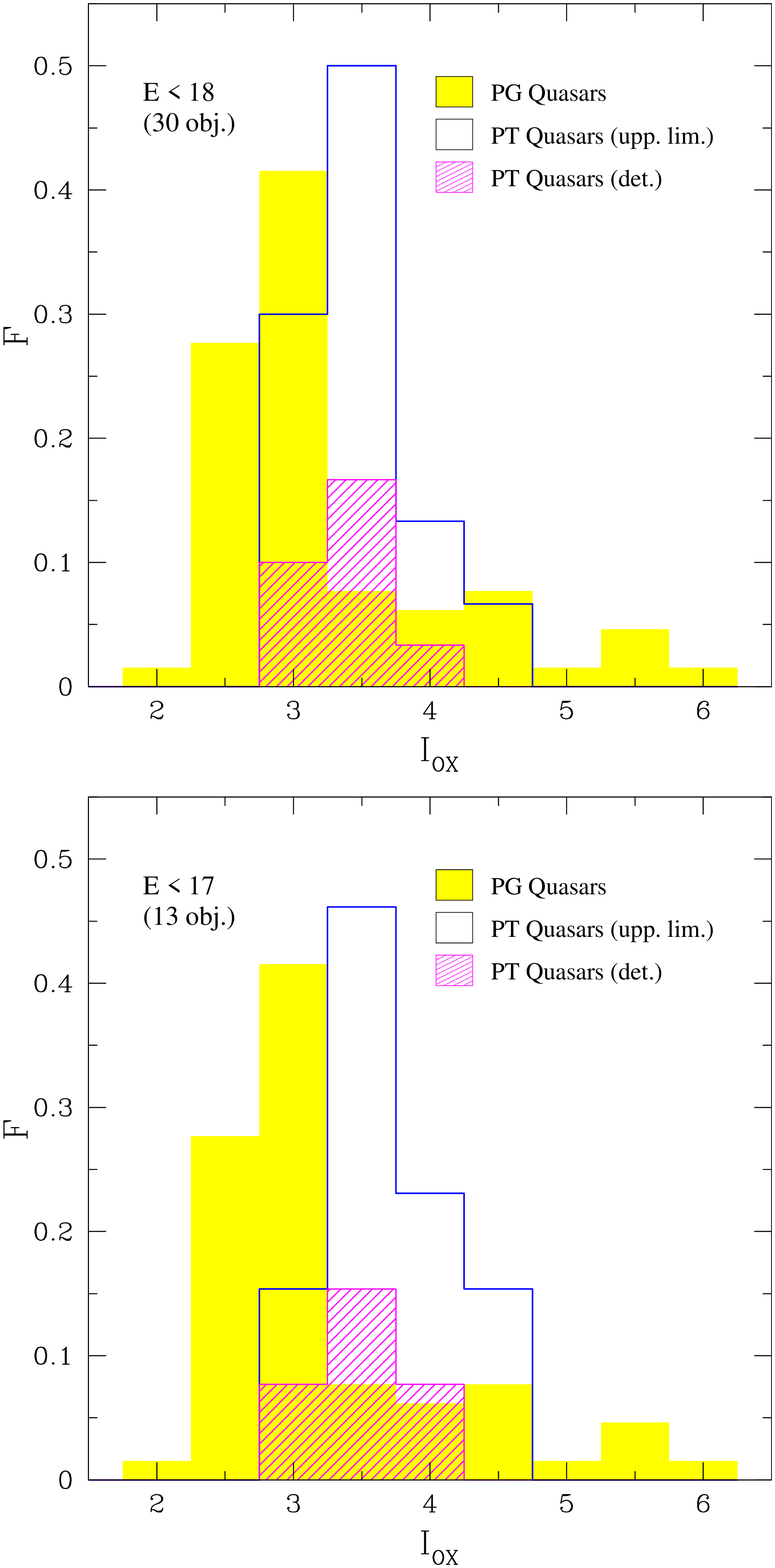}}}
\caption{\label{fig:pthiox}
\footnotesize{I$_{OX}$ distribution for the objects of the PT
sample with E $<$ 18 (upper panel) and with E $<$ 17 (lower panel).}}
\end{figure}

The I$_{OX}$ distribution for the PT sources is plotted in
 Fig. \ref{fig:pthiox}
for two diffent cuts in magnitude. A highly significant
difference between our sample and the PG quasars is evident even in the
E $<$ 18 sample: more than half of the objects are underluminous
in the X-rays of at least a factor 5 with respect to a ``normal''
quasar. The X-ray weakness is however more significant
in the small E$<$ 17 subsample. The increase of the discrepancy
with brightness shows that grism selection is indeed effective
in finding out X-ray weak quasars, but in the case of the PT sample
the flux limits in the WGACAT are too high to investigate the correlation
between the optical and X-ray emission.

\section{Selection effects and interpretation}

The results presented above suggest the existence of a class of AGNs
with broad lines in the
optical/UV but with an X-ray emission much lower than expected from the
optical flux and a PG-based extrapolation.

\begin{figure}
\centerline{\resizebox{\hsize}{!}
{\includegraphics{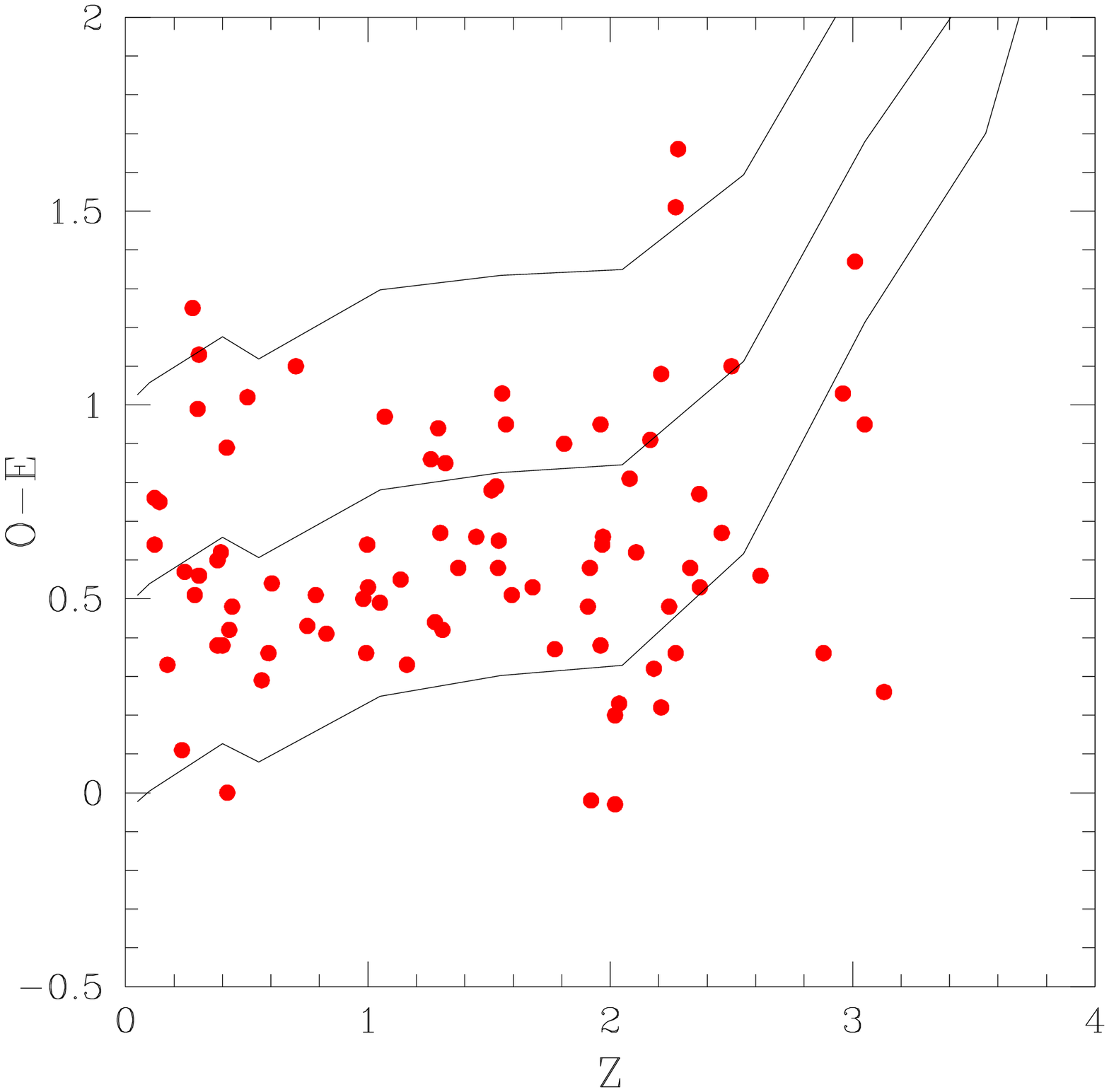}}}
\caption{\label{fig:hsoptx}
\footnotesize{Optical O-E
colour versus redshift for the HS sample. Lines are for a standard quasar spectrum with three
different rest frame optical extinctions: A$_V=0$ (bottom line), A$_V=1$ (middle line),
A$_V=2$ (top line)}}
\end{figure}

\begin{figure}
\centerline{\resizebox{\hsize}{!}
{\includegraphics{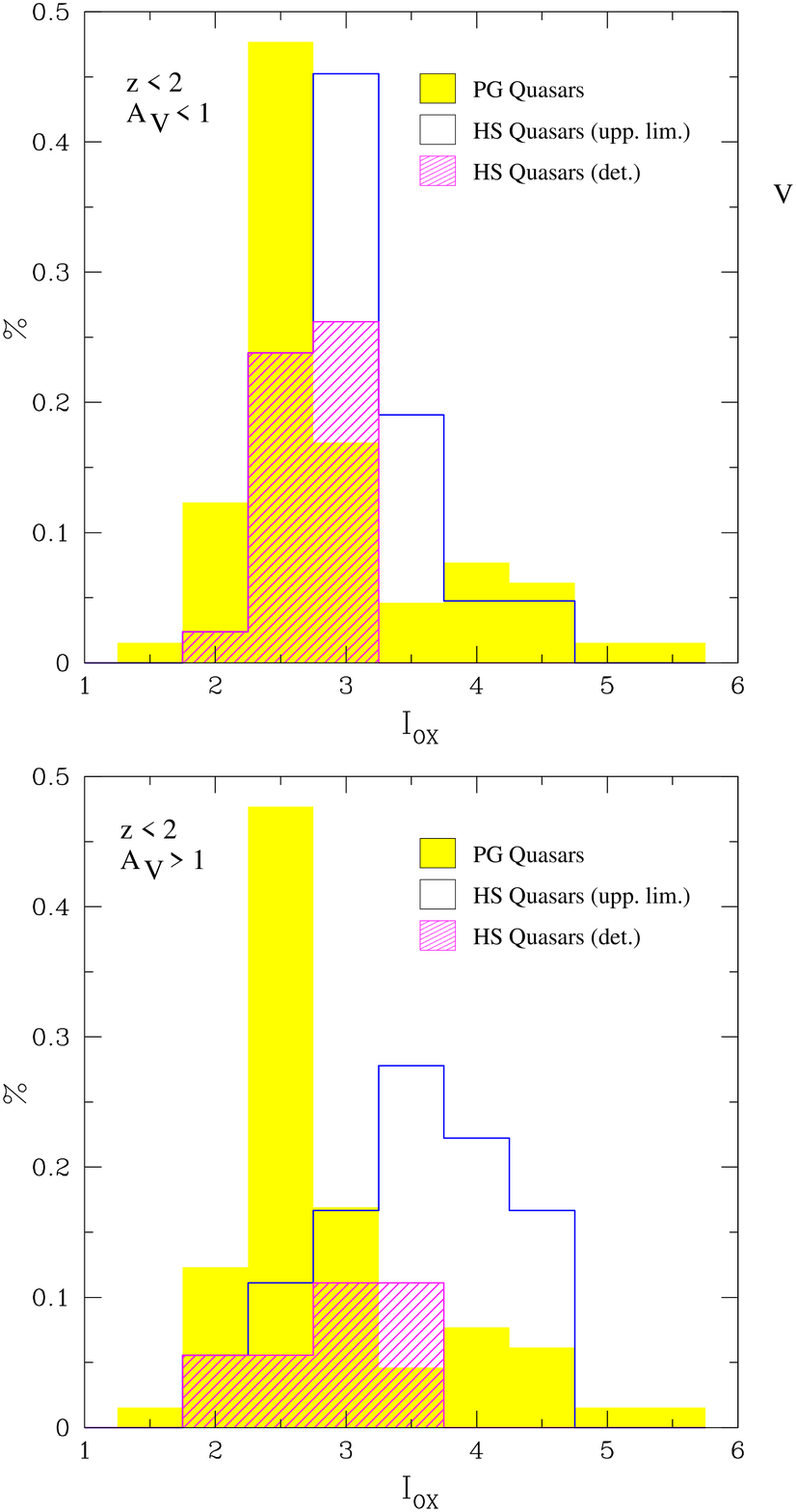}}}
\caption{\label{fig:hsoptx2}
\footnotesize{
I$_{OX}$ distribution for the z$<2$ HS sample below (upper panel) and above
(lower panel) the
A$_V=1$ line in Fig. 5. }}
\end{figure}

The quasars of the HS sample show a significant correlation
of their broad-band properties 
with redshift (or, analogously, with luminosity, since the sample is
flux--limited). 
We performed a KS test on the distributions plotted in Fig. \ref{fig:hsiox}, and we 
found that the z$<$1 and 1$<$z$<2$ distributions are different at a level of confidence of
96\%, while the z$<1$ and z$>2$ distributions differ at 99\%.

We note that the level of significance of the lower limits on I$_{OX}$ is the same
at all redshifts, because both the optical and the X--ray parent samples are flux
limited at {\it constant} fluxes.
The increase of I$_{OX}$
with redshift can therefore be due to a ``luminosity effect'': optically
more luminous quasars are on
average weaker in the soft X-rays (with respect to the optical emission).

In Fig. \ref{fig:hsoptx}
 we show the O-E colour for the HS sample, obtained from of the
Palomar All Sky Survey (POSS). The theoretical lines represent the typical O-E
colour of quasars in function of the redshift, for three different values of
optical extinction. Since the blue colour excess is still one of
the main selection
criteria here, more than half of the HS quasars do not have, on
average, colours redder than normal quasars.
However,
another $\sim$25\% of the sources have an O-E colour that corresponds in normal
quasars to an optical extinction A$_V > 1$. This is a further indication that
also quasars with a relevant absorption were selected in the Hamburg Quasar
Survey. 
In Fig. \ref{fig:hsoptx2} we show the I$_{OX}$ distributions for
the HS objects with z$<2$ and located respectively below (upper panel) and above (lower panel)
the A$_V$=1 line in Fig.\ref{fig:hsoptx}.
Redder objects are clearly also on average weaker in the
X-rays. 
This results confirm one of our main working hypotheses, i.e. that X-ray weak quasars are
missed in surveys based on a strong colour selection.

Sources at z $>2$ are significantly X-ray weaker than normal quasars, irrespectively
of their optical colour. This is in agreement with the ``luminosity effect''
discussed above. Again, this class of objects could easily have been missed
by colour-based surveys, since the U-B criterion becomes inefficient at z$>$2.2.

Since the HS survey is still biased towards blue objects, this result could imply that a non negligible 
population of ``red'' quasars could exist, analogously to what found by Webster et al. (1995)
for flat-spectrum radio quasars.

\begin{figure}
\centerline{\resizebox{\hsize}{!}
{\includegraphics{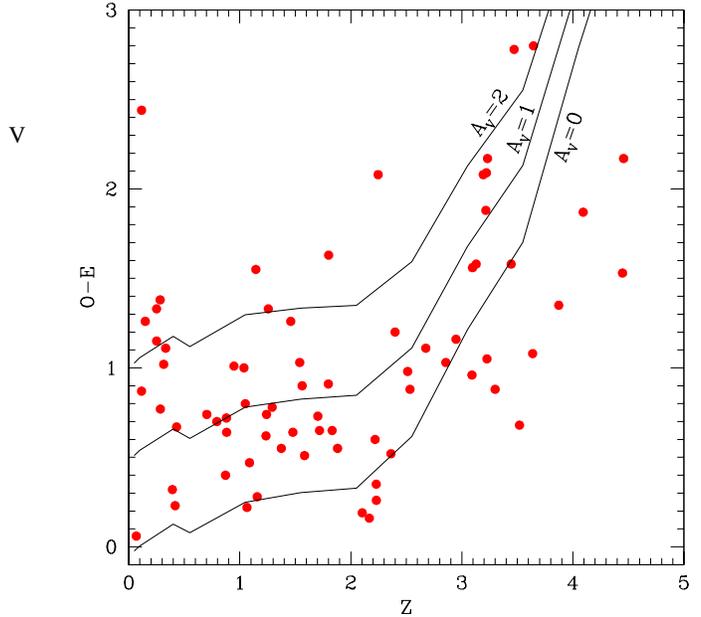}}}
\caption{\label{fig:pthoptx}
\footnotesize{
O-E colour for the PT quasars . The lines are the
same as in Fig. 5.}}
\end{figure}

The discussion made above for the HS quasars can be repeated for the 
PT quasars. In this case, even if we are able to find 
a significant fraction of X-ray weak quasars, the result is weaker,
due to the fainter optical magnitudes of these sources.

The O-E colour distribution for the PT sample is similar
to that of the HS, revealing a significant fraction of objects redder than
normal quasars (Fig. \ref{fig:pthoptx}). This result confirms that spectroscopic
selection is able to find quasars that would be missed in classical color-based
surveys. 

\section{Discussion and conclusions}

The X--ray and optical analysis of the HS and PT samples,
presented in the previous Sections, points out
the existence of a population of X--ray weak AGNs, with optical colours
not very different from ``normal'' quasars, but red enough to be missed by
classical colour--based surveys.

Our results on the I$_{OX}$ distribution imply that at least 50\% of our objects are
significantly weaker in the soft X rays than normal ``blue'' quasars. 

The high I$_{OX}$ measured in our sources could be due to an 
intrinsic
X-ray weakness, or to a low A$_V/$N$_H$ value. Several considerations make the first hypothesis
unlikely:
\begin{itemize}
\item Soft X-ray emission is a general feature present in all known classes of AGNs (except
for BALs, but see the next paragraph). 
Moreover, since the strong emission lines detected in
the optical require a powerful UV source, it is unlikely that a physical process provides
an intense UV emission without any significant tail in the soft X rays. 
\item As briefly mentioned in the Introduction, BAL AGNs, that are the only sources
similar to ours with respect to the X/optical properties, are likely to be 
heavily absorbed rather than
intrinsically weak. This is suggested (a) by the correlation between the X to
optical ratio and the C IV absorption features (Brandt et al. 2000) 
and (b) by detailed analysis of
the X--ray emission of some BAL quasars, that shows an absorbed spectrum in
the 2-10 keV band (Gallagher et al. 1999). 
\end{itemize}

On the other hand, there are several indications that a low A$_V/$N$_H$ (relative to the
Galactic value) could be common to several classes of sources (see the Introduction).
Moreover, this possibility is supported by the correlation between high I$_{OX}$ values and
high O-E colours, suggesting a common physical origin of the two phenomena.

Current models of the circumnuclear medium of AGNs do not clarify 
which physical process can be responsible for an A$_V/$N$_H$ value substantially lower than
Galactic. Qualitatively,
a high absorption in the X rays, together with a low extinction in the
optical can be due to two different
reasons: a low dust--to--gas ratio or a size distribution of dust grains
different from the standard one. 
For a more detailed discussion on this issue, the reader is referred
to Maiolino et al. 2001b.

\begin{figure}
\centerline{\resizebox{\hsize}{!}
{\includegraphics{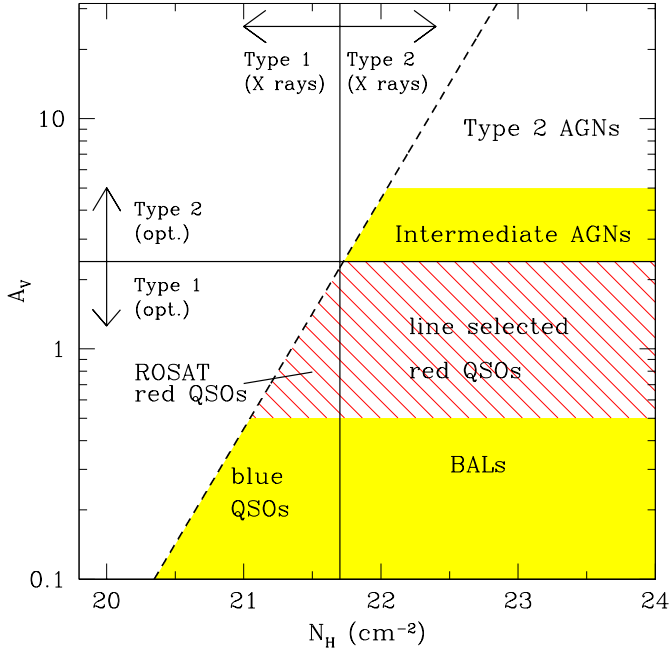}}}
\label{schema}
\caption{\label{fig:schema}
\footnotesize{AGN classification in the A$_V-$N$_H$ plane. The dashed line
represents a Galactic A$_V/$N$_H$ value. See text for details.}} 
\end{figure}

Whatever the physical origin of the low A$_V/$N$_H$, the main conclusion of this
discussion is that a population of quasars does exist,
 with X-ray properties of type 2 AGNs
and optical spectrum of type 1s.
In Fig. \ref{fig:schema} we propose a simple classification scheme for AGNs in
the A$_V$ - N$_H$ plane. In this diagram, ``blue''  quasars and BALs are the
objects revealed by classic, colour-based optical surveys. The
line-selected objects are located in the ``X-ray type 2 zone'' (N$_H > $ a few
10$^{22}$ cm$^{-2}$) but in the ``optical type 1 zone'' (A$_V < 2-3$). ``ROSAT
red quasars'' refers to the soft X-ray selected objects of Kim \& Elvis (1998).
It is
worth noting that even in the ``optical type 2 zone'' a relevant number of
objects could have low A$_V/$N$_H$: for example, comparing the near IR broad
lines with the X-ray spectrum of the type 2 AGN IRAS 05189-2524, we estimated
an A$_V/$N$_H$ substantially lower than Galactic (Severgnini et al. 2000).

\begin{acknowledgements}
This work has made use of data obtained through the VizieR Service for
Astronomical Catalogues at CDS, Strasbourg, France, and the APM Catalogues
provided by the Institute of Astronomy, Cambridge, UK.

The authors acknowledge partial financial support 
by the Italian Space Agency (ASI) under grant
ARS-99-15 and by the Italian Ministry for University and Research (MURST) under
grant Cofin98-02-32.
\end{acknowledgements}

\appendix
\section{Evaluating upper limits from the WGACAT catalogue}

Here we discuss how to obtain a correct estimation of the upper limits (in counts s$^{-1}$)
for a non-detection in the WGACAT catalogue.

The basic selection criterion in the WGACAT is the request of a signal-to-noise ratio higher
than 2.
This implies that, if the background counts are B, the minimum number of 
counts for a detected source, $S_m$, is
obtained from the equation:

\begin{equation}
S_m = 2\sqrt{S_m+B}
\end{equation}

In case of a non-detection, a 90\% upper limit is obtained multiplying $S_m$ by a correction
factor {\em a}, such that a source with $aS_m$ counts has a probability $<$10\% to be undetected
(i.e. to fall below the detection limit $S_m$). This factor is therefore obtained solving
the equation:

\begin{equation}
\sigma \sqrt{aS_m+B}=S_m (a-1)
\end{equation}
where $\sigma=1.25$ is the number of standard deviations corrisponding to a 1-sided 
probability of 90\%.
We found that typical values for this correction are $a \sim 1.65-1.7$.

The crucial point in order to obtain a correct value of $aS_m$ is the estimation of B.

For each undetected source, we calculated the background counts for a nearby detected source
and we re-scaled this value for the vignetting snd PSF of the PSPC instrument.

A tabulation of the vignetting as function of the distance from the center of the field
of view, $y(r)$, can be obtained directly from the WGACAT.
To obtain the PSF we calculated the background counts B for a large number of randomly
selected sources (about 10\% of the WGACAT) and we fitted the dependence from the radius 
with a 3rd order polynomial, P(r). Finally we re-scaled the background counts of the nearby detected
source with the ratio $[P(r_1)y(r_1)]/[P(r_2)y(r_2)]$, where $r_1$ and $r_2$ are the distances
from the center of the undetected and the detected source, respectively.

\end{document}